# Confinement-Driven Exciton Behavior in 2D Halide Perovskites from Dielectric-Dependent Hybrid Methods

Rafael B. Araujo[a,*], Mustafa Mahmoud Aboulsaad[a], Sebastian E. Reyes-Lillo[b], Tomas Edvinsson[a,*]

[a] Department of Materials Science and Engineering, Solid State Physics, Uppsala University, Box 35, 75103 Uppsala, Sweden

[b] Departamento de Física y Astronomía, Universidad Andres Bello, Santiago 837-0136, Chile

[*] Corresponding authors: rafael.araujo@angstrom.uu.se, tomas.edvinsson@angstrom.uu.se

Abstract

Understanding how dielectric anisotropy governs excitonic behavior in two-dimensional (2D) halide perovskites is critical for predicting and engineering their optoelectronic properties. In this work, we investigate $Cs_{(n+1)}Pb_nBr_{3n+1}$ nanoplatelets (n = 2–5) experimentally and theoretically and show that the interplay between dielectric confinement and anisotropic screening critically determines both their electronic structure and excitonic landscape. To incorporate the dielectric screening effects, the Coulomb kernel in the Fock exchange term is refined using a model dielectric function together with a model Bethe-Salpeter Equation approach. The exciton binding energies show a monotonic decrease from 0.26 eV to 0.21 eV from n = 2 to 5, with 20 meV decrease per layer up to n = 4, and thereafter less change. The relatively small change per layer is a consequence of the strong spatial localization of excitons. By analyzing directionally resolved dielectric tensors, we demonstrate that the in-plane dielectric constant predominantly dictates optical transitions and is close to converging to the bulk value already at n = 5, while the out-of-plane dielectric response reflects the confined nature of excitonic wave functions as expected. Our calculated absorption spectra capture experimental results within 0.02 eV throughout the confinement regime (n = 2-5). The effects of lattice dynamics on the dimensionally dependent dielectric response and subsequent exciton screening occurring on longer time-scales than the optical response are also analyzed, important for analysis and interpretation of exciton lifetime, diffusion, and band alignments. The results establish a clear correlation between dielectric anisotropy, electronic structure, and exciton binding energy at different timescales in layered perovskites, providing essential insight for the design of 2D optoelectronic materials and devices.

Quantum confinement and reduced dimensionality profoundly reshape electronic and optical properties in semiconductors, particularly by enhancing Coulomb interactions and altering dielectric screening.[1] This effect happens when the size of a material is reduced to the nanoscale, limiting how electrons and holes can move. In two-dimensional (2D) materials, quantum confinement can increase the strength of light-matter interactions, raise the energy needed to separate electron-hole pairs (quantified with the exciton binding energy), and lower the ability of the material to screen electric fields.[2,3] The strength of these changes depends on the type of material.[4] 2D halide perovskite nanoplatelets (NPLs) have emerged as promising materials due to their tunable optoelectronic properties, driven by confinement and dielectric effects between layers and surroundings.[5,6]

Tuning the optical response in perovskites through compositional changes is a well-explored area of research. By changing or mixing halide ions and A-site cations, one can manipulate the bandgap and optoelectronic properties of these materials. For instance, incorporating different halides can lead to a range of absorption spectra, which is crucial for applications in photovoltaics and light-emitting devices.[5,7] More recently, the focus has shifted to the quantum confinement that naturally comes from the layer thickness of low-dimensional heterostructures, using it as a powerful way to control and tune the optical and electronic properties, and in an extension, how they function in an operating device. While most previous studies have explored 2D NPLs for practical and operational conditions [2–4,8], the fundamental role of dielectric anisotropy in governing exciton behavior across varying thicknesses remains insufficiently explored. In reduced-dimensional perovskites, spatially confined excitons experience directionally dependent screening, with in-plane and out-of-plane dielectric responses contributing asymmetrically to the binding energy and absorption spectra. Capturing this anisotropy is essential for the quantitative prediction of excitonic properties. Excitonic behavior cannot be effectively described using a single-quasiparticle approximation, such as that employed in density functional theory (DFT), turning the accurate prediction of excitonic effects into a central challenge. A widely used approach to compute quasiparticle states and for examining excitonic properties involves perturbation theory within the GW approximation, followed by solving the Bethe-Salpeter Equation (BSE) to determine optical spectra.[9] However, this method is usually computationally expensive for halide perovskites, which require spin-orbit interaction, and for large supercells and 2D systems, which require significant vacuum regions along the out-of-plane direction and large planar dimensions. There are, however, alternatives that maintain the accuracy of the GW-BSE approach keeping a reasonable computational effort. Notably, time-dependent density functional theory (TD-DFT) combined with non-empirical hybrid functionals has been shown to achieve results comparable to BSE while significantly reducing computational costs.[10,11] This approach has been employed to study self-trapped excitons and complex excitonic states in low-dimensional systems.[12–15] Another alternative is the model BSE (mBSE) approach, in which the GW-screened interaction is substituted by an empirical model.[10,16,17] This avoids the computational expense of GW calculations but comes at a cost: since the band structure is computed at the DFT level, the band gap values are likely underestimated.

In this study, we investigate the impact of anisotropic dielectric screening on excitonic properties by applying dielectric-dependent hybrid (DDH) functionals in combination with the mBSE framework to $Cs_{(n+1)}Pb_nBr_{3n+1}$ nanoplatelets (n = 2–5). This approach utilizes the directional components of the high-frequency dielectric tensor to consistently link electronic structure with optical response while maintaining reasonable computational demand.[18–20] As the layer thickness increases, the effects of quantum confinement diminish, resulting in a systematic reduction of the band gap. The exciton binding energy decreases monotonically from 0.26 eV to 0.21 eV with approximately 20 meV per monolayer. Although decreasing, the change in exciton binding energy remains relatively small, indicating strong spatial confinement of the excitonic wavefunction regardless of the number of layers up to n = 5. Our findings show that excitonic transitions are primarily governed by in-plane dielectric screening, highlighting the crucial role of anisotropic polarization in shaping optical selection rules and binding energies. The calculated absorption spectra, based on this dielectric model, align closely with our experimental observations, supporting the accuracy of the anisotropic framework. Overall, this work demonstrates how directional dielectric properties influence excitonic behavior in confined systems and presents a transferable strategy for designing 2D optoelectronic materials across a wide range of anisotropic semiconductors. In addition to predictive modeling, our results provide practical design guidelines for engineering perovskite-based devices—such as LEDs, solar cells, and detectors—with tailored excitonic responses.

To carry out our investigation, we adopted a multistep computational strategy that integrates DDH functionals with mBSE. While DDH combined with time-dependent DDH (TD-DDH) has been used in prior studies to estimate excitonic effects, this approach presents significant challenges, particularly the need for extremely fine k-point sampling to accurately capture exciton binding energies. To overcome this limitation, we propose coupling the DDH framework with the mBSE formalism. Our workflow read as:

1. Compute band structures using the DDH functional.
2. Apply an operator to the DFT wavefunctions to propagate the DDH band gap into the model BSE framework. Here, a much finer k-grid is employed to obtain the absorption spectra.
3. Ensure consistency by using the same inverse high-frequency dielectric constant in the mBSE approach as that used in the DDH calculations.

Unlike conventional methods that rely on a fixed fraction of exact exchange (e.g., 25% in PBE0[21] or HSE06[22]), the DDH functional determines the fraction of exact exchange dynamically based on the material's dielectric properties, particularly the inverse of the high-frequency dielectric constant.[18,19] This adaptability enables DDH functionals to provide a more accurate and universal description of band structures. To incorporate the material's dielectric screening effects, the Coulomb kernel in the Fock exchange term is refined using a model dielectric function.[11]

$$\epsilon^{-1}(q + G) = 1 - \left(1 - \frac{1}{\varepsilon_\infty}\right) exp\left(-\frac{|q+G|^2}{4\mu^2}\right) \quad (1)$$

In this formulation, $\mu$ represents the range-separation parameter. The term $\frac{1}{\varepsilon_\infty}$, quantifies the material's long-range screening at optical frequencies. This modification ensures that short-range Coulomb interactions remain unscreened, capturing local exchange effects accurately, while long-range interactions are effectively screened based on the material's dielectric environment. Compared to traditional hybrid functionals, DDH functionals provide a more accurate description of band gaps and electronic properties without requiring empirical fitting.[18–20]

The n = infty end-member of the Cs$_{(n+1)}$Pb$_n$Br$_{3n+1}$ system CsPbBr$_3$ typically crystallizes in four distinct structural phases in its bulk form: a high-temperature cubic (α-phase), an intermediate tetragonal (β-phase), and two orthorhombic configurations—one being the conventional γ-phase and the other a non-perovskite X-phase. As the temperature rises, CsPbBr$_3$ undergoes structural transitions: it transforms from orthorhombic to tetragonal around 380 K, and then transitions to the cubic phase at approximately 403 K.[23] Across these phases, the Pb$^{2+}$ and Br$^-$ ions form corner-sharing PbBr$_6$ octahedra. However, the orientation and connectivity of the octahedra vary with the structure, and the phase transitions are driven by phonon-induced octahedral tilting.[24] Notably, it has been shown that these ideal tetragonal and cubic phases do not exist in a perfectly ordered manner at the local scale. Instead, due to inherent static and dynamic lattice disorder, the material exhibits local structural fluctuations, making its apparent symmetry tetragonal or cubic only when viewed over larger length scales.[25] Here, we follow our investigation using an orthorhombic structure that is the bulk ground state at room temperature.

The first step in performing the DDH calculations for Cs$_{(n+1)}$Pb$_n$Br$_{3n+1}$ (where n = 2, 3, 4 and 5) is to determine the appropriate values of μ. This is achieved by fitting Eq. 1 to the RPA dielectric function obtained from GW calculations for bulk CsPbBr$_3$ in the orthorhombic phase.[11] The results, shown in Figure 1a, indicate an optimal μ value of 0.95, corresponding to a high-frequency dielectric function of 3.5. However, this method cannot be directly applied to 2D slabs due to the computational cost of the RPA approach, which becomes significantly demanding for systems with large vacuum regions, as these require a higher number of plane waves. Therefore, for the 2D structures, the values of μ determined from the bulk phase will be used, while the value of the high-frequency dielectric function will be estimated with the following alternative approach.

One solution is to extract the high-frequency dielectric function of the investigated materials using the HSE06 functional. However, for computational systems such as the 2D slabs used in this study, additional considerations must be taken into account due to their reduced periodicity. In these cases, the three-dimensional periodic boundary conditions commonly applied in crystalline bulk calculations are artificially enforced by embedding the 2D system within a three-dimensional unit cell. A vacuum region is introduced to spatially separate the slab from its periodic replicas along the non-periodic direction. To extract key properties for a 2D slab system, the capacitor stack model can be employed.[26–28] The dielectric constant of the supercell is expressed as:

$$c\varepsilon_\parallel^{sc} = D\varepsilon_\parallel^{vc} + d\varepsilon_\parallel^{slab} \qquad (2)$$

$$c\varepsilon_{\perp,sc}^{-1} = D\varepsilon_{\perp,vc}^{-1} + d\varepsilon_{\perp,slab}^{-1} \qquad (3)$$

Here, *sc* refers to the supercell, *vc* to vacuum, and *slab* stands for the 2D layer. The labels $\parallel$ $and$ $\perp$ refer to the dielectric constants in the direction parallel and perpendicular to the layer, respectively. The values of $\varepsilon_\parallel^{sc}$ and $\varepsilon_\perp^{sc}$ are obtained employing finite differences with a perturbing electric field on the supercell, while the values $\varepsilon_\parallel^{slab}$ and $\varepsilon_\perp^{slab}$ are obtained using Eqs. (2) and (3). Finally, c is the cell size in z direction, D is the vacuum size in z direction and, here, d is the slab width. The width of the 2D slab is calculated as the difference between the atomic elements with higher z values and lower z values. Details are provided in Table S1.

To construct the 2D models used in this study, the orthorhombic bulk structures of CsPbBr$_3$ were first optimized. These refined structures were then used to generate slabs with varying thicknesses: two, three, four, and five layers (Figure 1a, bulk structural parameters summarized in the SI and Figure S1). To minimize interactions between periodic images and simulate the impact of organic components in NPL samples, a vacuum layer of at least 20 Å was introduced along the c-axis (001). The terminations of the 2D layers were carefully chosen to prevent dangling bonds, thereby ensuring structural stability. Moreover, we tested for the case of n = 2, building the slab in other directions such as (101). However, this orientation is about 0.3 eV higher in energy per formula unit than (001). Furthermore, to enhance the reliability of the structural configurations, a brief iterative ab initio molecular dynamics (IAMD) simulation was performed before the energy minimization step, where lattice parameters remained fixed. This additional step helps the system escape from local energy minima, mitigating structural bias and leading to a more physically accurate representation of the octahedral distortions typical in perovskites – partially introducing local disorder in the structure. Here, it is also important to highlight that the orthorhombic phase of the CsPbBr$_3$ is already a low-symmetry structure that considers local PbBr$_6$ distortions (2D structures used here are reported in the SI).

The n = 1 configuration is a valuable limiting case but was excluded here for two reasons. First, under our synthesis conditions, the inorganic CsPbBr$_3$ monolayer is not structurally stable without major changes in surface chemistry[29], whereas thickness-defined platelets from 2–5 ML are reproducible and structurally well-behaved. Second, our DDH-mBSE workflow is designed to quantify intrinsic anisotropic electronic screening of the inorganic slab across a constant chemistry/termination set. At n = 1, the dielectric response is dominated by surfaces and the external environment, making slab thickness and dielectric functions ill-defined within the capacitor model and breaking comparability with n = 2–5. A rigorous treatment of n = 1 would require explicit ligands/substrate in the dielectric model. Hence, we focus on n = 2–5, which represent structurally stable, experimentally accessible nanoplatelets.

Calculations were carried out utilizing the Projected Augmented Wave (PAW) approach within the Vienna Ab initio Simulation Package (VASP).[30,31] The generalized gradient approximation of Perdew, Burke, and Ernzerhof (PBE) was used to address the exchange and correlation terms within the Kohn-Sham Hamiltonian.[32] Structural optimization was performed until forces were below the threshold value of 0.01 eV/Å. Plane waves were expanded to a cutoff of 400 eV, and the Brillouin zone was sampled with a 3x3x1 grid for the HSE06 and DDH calculations and a 7x7x1 grid for the mBSE calculations. Dielectric constant tensors were computed from the derivative of the polarizability, where the derivative is computed using finite differences. Direction and intensity of the perturbing electric field are the default values in VASP, considering dipole corrections. All our calculations explicitly include spin-orbit coupling (SOC). Input parameters for the main calculations are shown in the SI (INCARs for VASP calculations).

*Table 1: In-plane ($\varepsilon_\parallel^{slab}$) and out-of-plane ($\varepsilon_\perp^{slab}$) computed dielectric constants and HSE06+SOC calculated band gap (Eg) for n = 2-5 and bulk Cs$_{(n+1)}$Pb$_n$Br$_{3n+1}$.*

| n | $\varepsilon_\parallel^{slab}$ | $\varepsilon_\perp^{slab}$ | Eg (eV) |
|---|---|---|---|
| 2 | 4.22 | 2.22 | 2.18 |
| 3 | 4.11 | 2.22 | 2.10 |
| 4 | 4.09 | 2.39 | 2.00 |
| 5 | 4.05 | 2.46 | 1.94 |
| bulk | 4.00 | 3.92 | 1.77 |

Our results show that the in-plane high-frequency dielectric function decreases from 4.22 to 4.05 as n in $Cs_{(n+1)}Pb_nBr_{3n+1}$ increase from n = 2 to n = 5 (Table 1). Interestingly, these values are close to converge (99%) to the bulk value n = 5, where the in-plane direction (along a and b axes) yields a high-frequency dielectric function value of 4.00. On the other hand, the out-of-plane high-frequency dielectric function for the 2D materials increases as expected, but is significantly smaller for n = 5 and further from convergence to the bulk value (63%). This is because the polarizability of the electron cloud in the out-of-plane direction is still limited at n = 5 due to confinement effects. As the 2D layer width increases, confinement effects become weaker, resulting in higher dielectric constants for systems with more layers (higher n values).

To calculate band gaps using the DDH approach, it is necessary to assign the appropriate high-frequency dielectric function value ($\varepsilon_\infty$) in Eq. (1). However, this becomes challenging in anisotropic systems, where the dielectric function tensor components differ significantly. This issue becomes evident when comparing the high-frequency dielectric functions in the out-of-plane and in-plane directions of the 2D systems. The values show strong discrepancies along the in-plane and out-of-plane, and using any of these values as input for the DDH calculation would lead to completely artificial results. Specifically, using the out-of-plane dielectric function reduces the screening of the Coulomb interaction, resulting in stronger electron-electron interactions and higher band gaps. In contrast, using the in-plane dielectric function enhances Coulomb screening, leading to lower band gaps. This indicates that the transition from the top of the valence band to the bottom of the conduction band is primarily governed by the in-plane dielectric constant. Therefore, the calculations of the band gaps performed in the DDH framework will consider the in-plane dielectric constant of the 2D materials. Cui *et. al.*[18] have benchmarked the performance of the DDH and Doubly Screened Hybrid Functional (DSH) for different band gap semiconductors, showing systematic good performance. Interestingly, this work uses the maximum value of the diagonal of the high-frequency dielectric tensor as the intrinsic parameter in Eq. (1). So, the concept of using the higher value in the trace of the dielectric tensor is not new and has previously shown good results in predicting band gaps.

To assess the validity of our approach, we compare the experimental absorption spectrum of $Cs_{(n+1)}Pb_nBr_{3n+1}$ with *n* = 2 to the theoretical results presented here (experimental details are in the SI). To assess the uniformity and thickness of the here synthesized nanocrystal, we performed annular transmission electron microscopy (TEM) analysis (Figure S2 (a)). The reduced contrast of the side view indicates the formation of ultrathin nanoplatelets, showing a measured thickness of 1.2 ± 0.1 nm. Considering that a single monolayer corresponds to one 2D sheet of corner-sharing $[PbBr_6]_4^-$ octahedra with a thickness of approximately 0.6 nm, this value suggests that the nanoplatelets consist of two monolayers. To further confirm these structural characteristics, we recorded steady-state photoluminescence (PL) spectra for the same samples (Figure S2 (b)). The PL spectra display a single, well-defined and narrow emission peak centered at 431 nm, in excellent agreement with the results reported by Bohn et. al.[29], where 2 ML nanoplatelets exhibited a PL maximum at 432 nm.

We focus on three main computational cases for the experimental vs computed comparison. First, the spectrum is calculated using TD-DFT in combination with the DDH functional on a 3×3×1 k-point grid. Second and third, the spectra are computed using the mBSE approach with k-point grids of 3×3×1 and 7×7×1, respectively. This comparison allows us to examine the differences between TD-DFT and mBSE results on the same grid, as well as the effect of increasing grid density within the mBSE framework (convergence tests are displayed in Figure S3). It is important to note that, due to computational limitations, DDH calculations could not be performed on grids denser than 3×3×1.

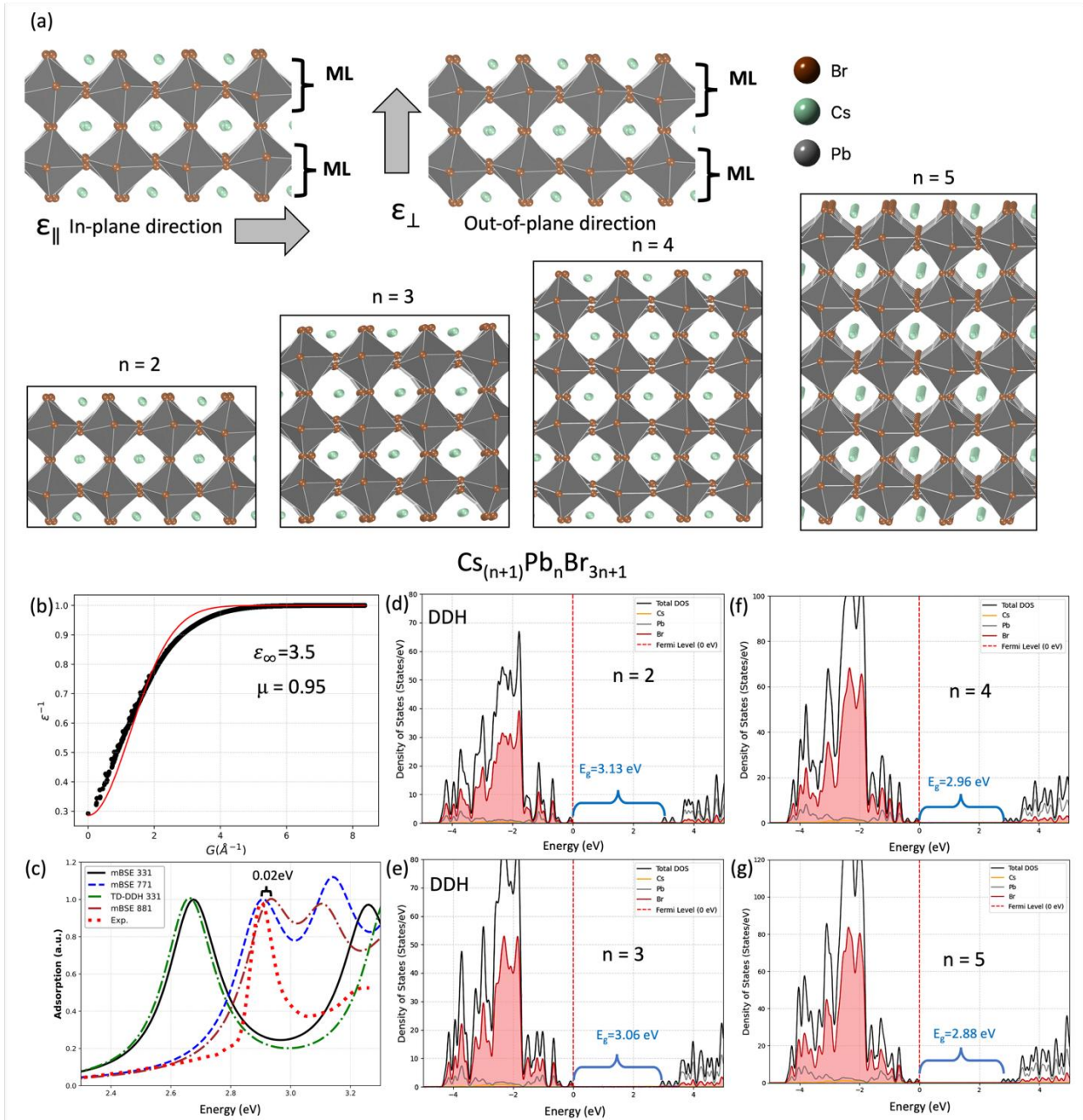

*Figure 1: a) Structural representation of the Cs$_{(n+1)}$Pb$_n$Br$_{3n+1}$ nanoplatelets (n = 2–5). b) Inverse dielectric function vs. wave vector at $\Gamma$ point for the bulk orthorhombic CsPbBr$_3$ and fitted Eq. (1). c) The absorption spectra obtained from mBSE and TD-DDH are compared with experimental data. The theoretical spectra were renormalized to align with the intensity of the first experimental peak. (d-g) Projected density of states on the atoms of the Cs$_{(n+1)}$Pb$_n$Br$_{3n+1}$.*

The differences between the computed absorption spectra for the n = 2 case using TD-DDH and the mBSE with a 3×3×1 k-point grid are relatively small (Figure 1c), resulting in similar overall features. However, the low k-point density leads to an erroneous estimation of the exciton bound-state energies. This results in an absorption peak around 2.7 eV, which is not observed experimentally. In contrast, when the k-point sampling is increased, 7×7×1, the mBSE-calculated absorption spectrum shows the first peak at nearly the same position as in the experimental data, around 2.90 eV. The second peak, however, remains overestimated compared to the experiment. As shown in Figure 1c, increasing the k-point density from 7×7×1 to 8×8×1 reduces the relative intensity of the higher-energy feature, while the first excitonic peak remains nearly unchanged in energy. This behavior arises because a denser k-mesh provides a more accurate description of the exciton binding energy and redistributes the oscillator strength from the continuum region toward the first excitonic transition. Consequently, the excessive intensity of the second peak observed at coarser grids is a convergence artifact rather than a deficiency of the model. Despite this overestimation, the good agreement of the first peak indicates that the calculated band gap is consistent with the experimental value and, importantly, that the first excitonic bound-state energy is accurately estimated - here for a convergence factor of 0.02 eV as shown in Figure 1c.

We emphasize that, although the calculated value of the first excitonic bound state agrees well with experimental data for n = 2, with a difference of less than 0.02 eV, this apparent accuracy is likely due to the strong confinement and clear distinction of the larger in-plane and smaller out-of-plane dielectric constant where any small variation in orientation of the nano platelets would matter in the experimental response. Here, one cannot rule out effects from error cancellation and k-grid convergence (here converged to 0.02 eV and therefore adding such uncertainty in the full value). Such cancellation may arise from over- or underestimation of the band gap, structural discrepancies between the experimental sample and our model, or other approximations here assumed, like surface terminations and k-point grid density. While this issue is noteworthy, the discrepancy between the band gaps obtained using the DDH functional and experimental values is not expected to be significant under typical conditions. Recently, Garba et al.[33] investigated the accuracy of the DSH functionals in predicting the band gaps of layered perovskites. They demonstrated that, when using the dielectric constant of the corresponding three-dimensional (3D) perovskite, the method yields band gap predictions for layered perovskites with a mean absolute error of 0.14 eV. In any case, residual inaccuracies would manifest as a rigid energy shift affecting all thicknesses similarly, without altering the relative spectral trends that are the focus of this work.

Here, we primarily employed the in-plane component of the high-frequency dielectric constant ($\varepsilon_{slab,\parallel}$) in Eq. (1), based on the assumption that excitonic transitions are predominantly governed by in-plane electronic polarization. However, as the slab thickness increases, the contribution from the out-of-plane dielectric response may become non-negligible. To evaluate this effect, we computed the absorption spectra for the n = 2 and n = 5 cases using three scenarios in Eq. (1): $\varepsilon_\parallel$, $\varepsilon_\perp$, and the averaged dielectric constant defined as $(2*\varepsilon_\parallel + \varepsilon_\perp)/3$ (Figure S4 – for the case of n = 2 and Table S2). While the main peak, associated with the first bright exciton, aligns best with experimental data when $\varepsilon_\parallel$ is used, employing either the averaged or $\varepsilon_\perp$ values for the high-frequency dielectric constant results in a relative lowering of the second peak compared to the first, consistent with the experimental spectral shape for the n = 2 case. However, in the case of the averaged dielectric constant, the position of the first peak is offset from the experimental value by approximately 0.19 eV, while for the second case, it is shifted by 0.78 eV. These findings support our choice of using $\varepsilon_\parallel$ in the excitonic model, as it more accurately reproduces the absolute position of the first bright exciton. We note that the ratio between the two peaks is better reproduced by using, for instance, the averaged high-frequency dielectric constant. The excitonic binding energies computed for the $\varepsilon_\parallel$, $(2*\varepsilon_\parallel + \varepsilon_\perp)/3$ and $\varepsilon_\perp$, are 0.26 eV, 0.29 eV, and 0.56 eV for the 2ML case. As expected, lower dielectric constants lead to higher binding energies. While this difference is significant for the case of $\varepsilon_\perp$ generating a much higher binding energy, using the averaged and the in-plane results in an energy difference of only 0.05 eV and keeping the in-plane dielectric constant maintains the physical consistency that the first excitation is dictated by the in-plane polarization. Similar results are observed for the 5ML case.

The partial density of states (PDOS) of the NPLs computed with the DDH-SOC level of theory and using the in-plane high-frequency dielectric function (Figure 1d-g) reveals a band gap with values decreasing from 3.13 eV for n = 2 to 2.88 eV for n = 5 (Table 2). The observed behavior is consistent with quantum confinement effects, where reduced confinement decreases the band gaps. This aligns with findings of Akkerman et al.[34], who demonstrated a decrease in band gaps for 2D NPLs, though using a model based on the cubic phase of the $CsPbBr_3$. Moreover, the PBE computed band gap for the n =2 case, 1.48 eV, is close to the reported value of Cucco et al.[35], 1.56 eV. The PDOS and band structure indicate that the top of the valence band is predominantly formed by Br states, albeit with some contribution from Pb states. Conversely, the bottom of the conduction band is mainly composed of Pb states, exhibiting some contribution from Br states. The PDOS shows contributions from states of both bromine and Pb in the energy range from 0 eV to -2 eV, but with a much stronger contribution from Br states. This indicates the high ionic bond character with electron donation from the Pb atoms to the Br atoms.

Band gaps for our limiting cases (n = 2 and n = 5) were computed using different dielectric constants, the in-plane ($\varepsilon\parallel$), out-of-plane ($\varepsilon\perp$), and the averaged dielectric constant $(2\varepsilon\parallel + \varepsilon\perp)/3$—to assess their influence on the results. The outcomes are summarized in Table S2. As expected, higher band gaps were obtained when using the averaged and out-of-plane dielectric constants. For the averaged dielectric constant, the band gaps were shifted by more than 0.22 eV compared to those calculated with $\varepsilon\parallel$. In the case of $\varepsilon\perp$, the shifts were 1.08 eV for n = 2 and for n = 5. Clearly, band gaps increase faster than the excitonic binding energies, irrespective of dielectric constant values. Moreover, the gap variation is similar for the case of n = 2 or n = 5, suggesting that the gap value is mostly determined by the in-plane dielectric value.

*Table 2: DDH-SOC calculated first excitonic peak, band gap (Eg), binding energy of the first bright exciton (Eb) in eV, and the first excitonic peak from the experimental data reported by Bohn et al.[29] (a) for n=2-5, as well as our experimental data for n = 2.*

| n | First excitonic peak (eV) DDH | Eg (eV) DDH | $E_b$ in eV DDH | First excitonic peak (eV) Experiments |
|---|---|---|---|---|
| 2 | 2.87 | 3.13 | 0.26 | 2.87[a] / 2.88 (this work) |
| 3 | 2.83 | 3.06 | 0.24 | 2.81[a] |
| 4 | 2.74 | 2.96 | 0.22 | 2.74[a] |
| 5 | 2.67 | 2.88 | 0.21 | 2.68[a] |

We now present the calculated absorption spectra obtained using the mBSE approach as a function of layer thickness (Figure 2). For n = 2-5, the first bright excitonic peak appears at 2.87 eV, 2.83 eV, 2.74 eV, and 2.67 eV, respectively. In all cases, the true lowest-energy excitation is optically inactive (a dark exciton) and lies just below the first bright exciton. Here, in a more simplistic picture where SOC interactions are not accounted for, the first excitation would likely correspond to a single-triplet transition where the electron-hole pair transition is not optically allowed, followed by a singlet-singlet transition (bright exciton).

The energy difference between the first and second allowed optical transitions is approximately 0.26 eV, 0.24 eV, 0.22 eV, and 0.21 eV for n = 2–5, respectively. Since the second peaks are associated with transitions involving the conduction band minimum—i.e., the onset of the quasiparticle band gap—these energy differences reflect the exciton binding energies (Table 2). Notably, these values are relatively consistent across different thicknesses, with variations of only a few meV, indicating that the primary shift observed in the absorption spectra with increasing thickness arises predominantly from changes in the band gap. Hansen et. al.[36] have proposed an empirical relation between 2D perovskite band gaps and exciton binding energies as $E_b = 0.18 E_{gap} - 0.24$. By using this equation with the DDH band gaps of 3.13 eV for the n = 2 case, we obtain a binding energy of 0.32 eV, while for the n = 5 case, with a band gap of 2.88 eV, we obtain a binding energy of 0.28 eV. Hence, a systematic difference of about 0.06 eV is obtained

and this value is within the deviation reported by Hansen[36]

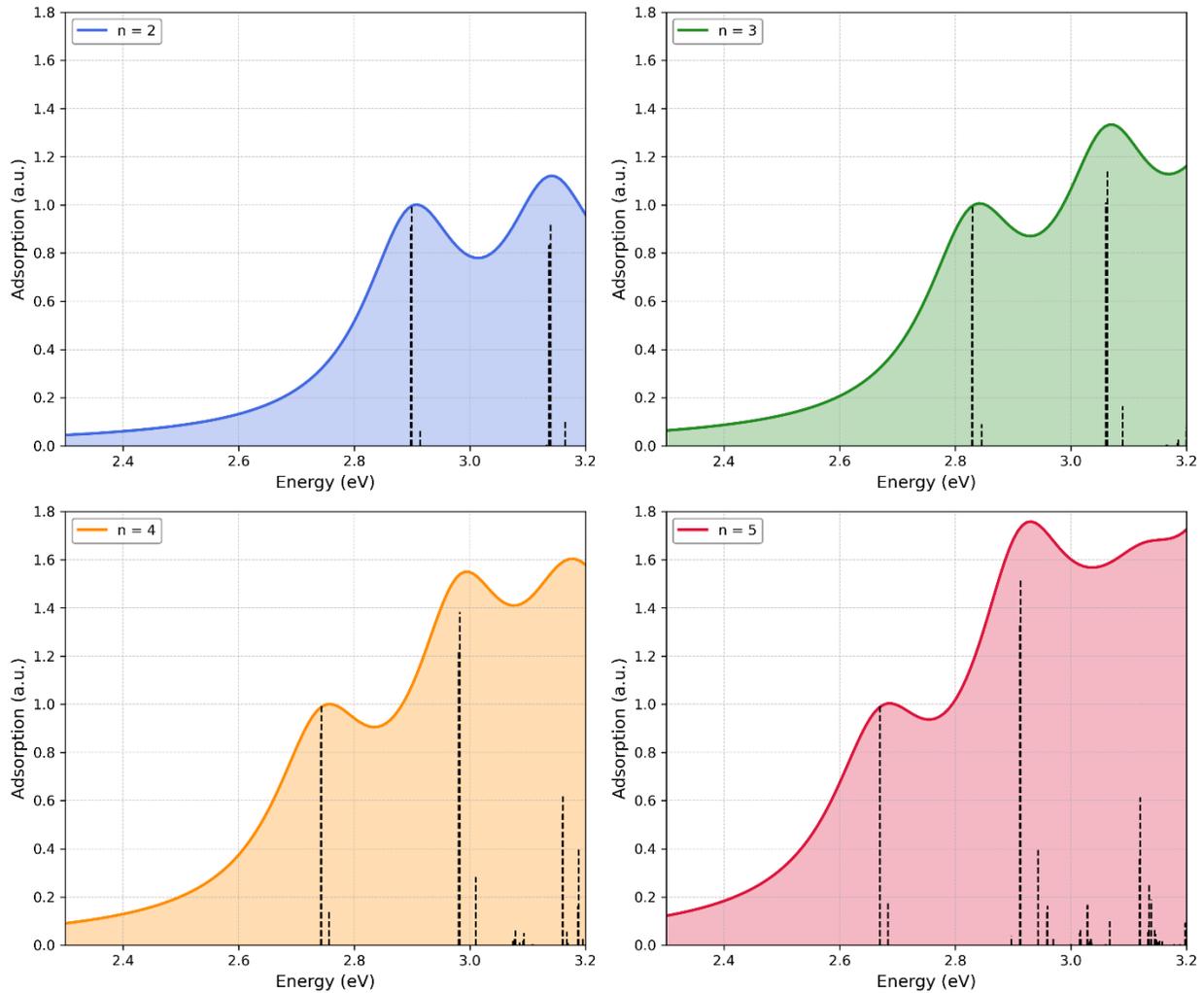

*Figure 2: Adsorption spectra computed at the mBSE level of theory for the $Cs_{(n+1)}Pb_nBr_{3n+1}$ with n = 2, 3, 4, and 5. Black vertical lines represent the probability of the transitions.*

The intensity differences between the first and second peaks observed in the absorption spectra are 0.1 a.u., 0.25 a.u., 0.55 a.u., and 0.78 a.u., respectively. Here, the intensities have been normalized such that the first peak is set to one. Although we focus on the first bright exciton peak, it is of interest to also look at the second excitonic peak. The increasing intensity of the second peak with layer thickness appears to be related to the higher density of states near the onset of the band gap. As the second peak becomes more pronounced, the absorption process in thicker slabs exhibits a reduced excitonic character and transitions toward a more band-like behavior. In the absorption spectra, this manifests as a diminished prominence of the excitonic peak. With increasing thickness, this excitonic feature gradually merges with the broader continuum of the spectrum. Due to this, a clear experimental signature of the second excitonic peak discriminated from band excitations is rare under ambient conditions, where for example a second distinguishable peak can only be seen for a 3 ML material in Bohn et al[29], in the range of 3.05-3.07 eV. The value is in good agreement with the second excitonic peak we found for n = 3 in Figure 2.

The nature of the first bright excitons for the n = 2 and n = 5 cases is further elucidated by the band-resolved contributions and the exciton charge density distributions (Figure 3). The dominant contribution to the first bright excitation arises at the Γ point, which corresponds to a relatively delocalized excitonic wavefunction in real space (generally, a more localized distribution in reciprocal space is associated with greater spatial delocalization of the exciton in real space and vice-versa). Notably, this behavior remains consistent between the n = 2 and n = 5 systems. Despite the significantly increased slab thickness in the n = 5 case, the excitonic charge density remains confined to approximately one of the five layers—similar to the distribution observed in the n = 2 structure (Figure 3). This spatial confinement is consistent with the comparable exciton binding energies obtained from the mBSE calculations, where excitonic binding energies are similar. A similar confinement effect has been shown in other perovskite systems.[37]

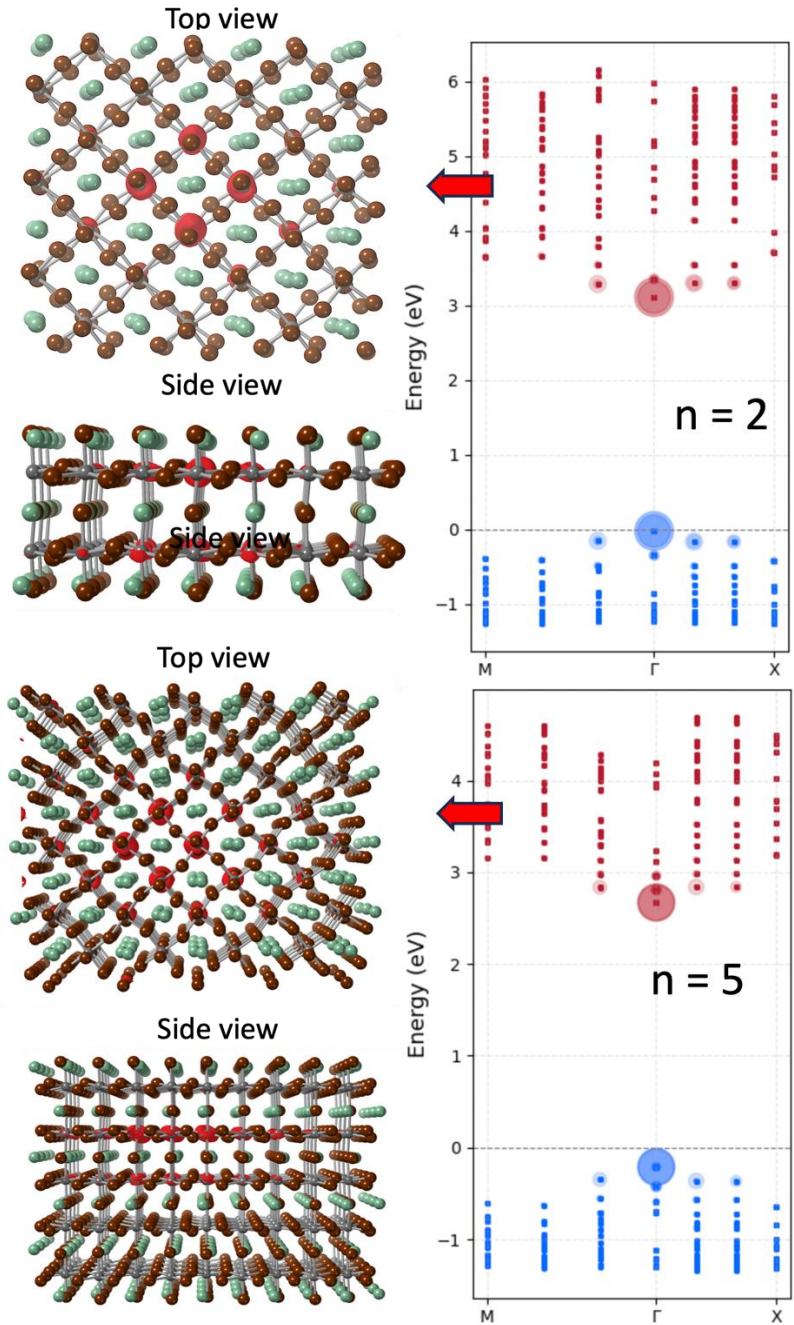

*Figure 3: (a) Top and side views of the exciton wave function of n = 2 and n = 5. The hole is fixed near a Br atom. The surface plot is produced using CrystalMaker software with 20% of the isosurface (b) Fatband band representation of the n = 2 and n = 5 slabs. Band eigenvalues are indicated by the black dots, with filled circles weighted by their contributions to the exciton. Hole and electron states are colored blue and orange. For the case of n = 2, one clearly sees the localization of the electron in the Pb atoms while holes are fixed in the Br atoms.*

To compare our results of the first bonded excitonic state energies with the Wannier–Mott model, we have calculated the carrier effective masses, $m^* = \hbar^2 \left[\frac{\partial^2 \varepsilon_n(k \rightarrow x)}{\partial (k \rightarrow x)}\right]^{-1}$, where $k \rightarrow x$ stands for the reciprocal components and $\varepsilon_n$ is the dispersion of the $n^{th}$ band (here the top of the valence band and bottom of the conduction band). The first bound state energy of the exciton is determined using the following expression:

$$E_b = \frac{\mu}{m_0} \frac{1}{\varepsilon_{slab,\|}^2} \frac{1}{\left(1 + \frac{\alpha - 3}{2}\right)^2} 13.6 \; eV, \quad \alpha = 3 - \gamma e^{-(L/2a_b)} \quad (4)$$

Where $\mu$ are the reduced effective masses of the electron-hole pair, and $\varepsilon_{slab,\|}$ is the high-frequency dielectric function of the slab in the in-plane direction (Table 3). This formula mixes a general empirical scaling law for exciton binding energies, grounded in a model tailored for low-dimensional systems, developed by Blancon et. al.[38], with adding effects of dielectric screening. Here $a_b = 4.6 \; nm$ is the Bohr radius of 3D perovskites[39], L is the width of the employed slabs for the different cases and $\gamma$ is an empirical correction factor. We also argue that only the $\|$ part of the 2D high-frequency dielectric constant has to be employed to derive the excitonic binding energy in the 2D Wannier–Mott model. Indeed, one of the assumptions behind the exciton model is the parabolicity of the band structure close to the bottom of the conduction band and the top of the valence band.[40] Due to the low dimensionality of the system along the z direction, the electronic bands should be flat or close to flat towards the associated reciprocal space direction, hence leading to a high or near infinite effective mass.[41]

Table 3: Width of the 2D slabs, electron and hole effective mass for the Γ-X directions, reduced masses (HSE06+SOC), calculated first bonded exciton binding energies with Eq. (4), calculated first bright exciton with the mBSE approach and calculated excitonic binding energy with the empirical model.

| n | L in nm | $m_h/m_0$ Γ-X | $m_e/m_0$ Γ-X | $\mu/m_0$ Γ-X | $\alpha$ | $E_b$ in eV, eq. 4, $\gamma = 1$ | $E_b$ in eV - mBSE | $E_b$ in eV, empirical model[36] |
|---|---|---|---|---|---|---|---|---|
| 2 | 1.25 | 0.30 | 0.30 | 0.15 | 2.13 | 0.36 | 0.26 | 0.32 |
| 3 | 1.83 | 0.31 | 0.31 | 0.15 | 2.18 | 0.36 | 0.24 | 0.31 |
| 4 | 2.44 | 0.30 | 0.30 | 0.15 | 2.23 | 0.32 | 0.22 | 0.29 |
| 5 | 2.99 | 0.29 | 0.29 | 0.15 | 2.28 | 0.29 | 0.21 | 0.27 |

The obtained exciton binding energies from Eq. (4) range from 0.36 eV to 0.29 eV for n = 2 - 5, when $\gamma = 1$, showing a general deviation of approximately 0.11 eV when compared to values obtained using the mBSE approach. Notably, the parameter α increases with increasing width, capturing the effects of quantum confinement. In contrast, the reduced mass, μ, remains nearly constant across all n values. Interestingly, and somewhat unexpectedly, the in-plane component of the high-frequency dielectric constant decreases with increasing width. While this change is minor, it could arise from the method used to determine the slab width in Eqs. (2) and (3), or from subtle structural variations—further investigation would be required to clarify this point. For the case where $\gamma = 0.7$, the obtained values of the exciton binding energies computed with Eq. (4), 0.24 eV, 0.24 eV, 0.23 eV and 0.22 eV for n = 2-5 are almost exact, the values from mBSE. This indicates that, though we added the effects of dielectric variation, there are still unaccounted factors in the Wannier–Mott expression.

We summarize our results by comparing them to the exciton binding energy obtained from the empirical model of Hansen et al.[36] and that obtained from the Wannier–Mott expression (Table 3). The small and systematic deviations across methods indicate that our framework consistently reproduces the confinement-dependent behavior observed experimentally, while emphasizing that empirical approaches can serve as quick estimation tools when full-scale BSE calculations are not feasible. Importantly, the agreement across all three methods—mBSE, modified Wannier–Mott, and empirical scaling—reinforces the robustness of the calculated trends and confirms the physical picture of confined excitons with moderate binding energies in 2D Cs-based perovskite nanoplatelets.

By comparing the band levels computed for the n = 2 structure using both the in-plane ($\varepsilon_\parallel$) and out-of-plane ($\varepsilon_\perp$) high-frequency dielectric constants (relative to the top of the valence band), we can clearly see why excitonic excitations are primarily governed by in-plane screening (Figure 4). Starting from the valence band maximum, the first optical transition appears at approximately 2.88 eV and corresponds to a bright exciton calculated using $\varepsilon_\parallel$—a result that aligns well with experimental data. The next feature, a band-like absorption edge, appears around 3.13 eV and is also associated with $\varepsilon_\parallel$. In contrast, the first bright excitonic state calculated with $\varepsilon_\perp$ appears much higher in energy, at about 3.65 eV. This state lies within the conduction band continuum and is therefore not observable in typical absorption experiments. While the exciton binding energy is larger when using $\varepsilon_\perp$—by approximately 0.30 eV—this is offset by a significant upward shift in the conduction band position when calculated with $\varepsilon_\perp$ instead of $\varepsilon_\parallel$. As a result, excitonic states associated with $\varepsilon_\perp$ appear at much higher energies. Ultimately, the optical absorption experiment only captures the excitonic transitions influenced by the in-plane dielectric response. These results underscore the importance of incorporating dielectric anisotropy and accurate electronic structure treatment to reliably capture excitonic properties in low-dimensional semiconductors. The same mechanism is obtained for the case of 5ML (Table S2)

It should be noted that the DDH–mBSE framework, as applied here, is based on ideal crystalline slabs and thus captures the intrinsic excitonic response in the absence of defects. In real perovskite systems, structural defects can introduce non-radiative channels that reduce photoluminescence efficiency and locally alter dielectric screening. Nevertheless, prior studies have shown that moderate defect densities often leave the fundamental band-edge excitons largely intact, consistent with the widely discussed "defect tolerance" of lead-halide perovskites.[42]

Substrate effects are also expected to influence excitonic properties through dielectric screening. In practice, these surfaces are passivated by ligands, which primarily stabilize the surface stoichiometry and introduce only minor local relaxation at the interface. Such relaxations are confined to the outermost atomic layers and do not cause the octahedral tilting or symmetry. A quantitative treatment of defect-mediated exciton behavior or substrate effects is beyond the scope of this work.

Exciton binding energies are usually derived from ultrafast optical measurements, where the lattice is effectively frozen and only the electronic screening contributes. However, at longer timescales, the lattice can respond via phonon polarization, introducing additional ionic screening that reduces the effective binding. If this effect is not properly accounted for, exciton binding energies can be misinterpreted. To better understand the role of dielectric screening and lattice dynamics on exciton formation in our nanoplatelets, we evaluated the exciton binding energy $E_b$ for the 2 ML and 5 ML cases using $\varepsilon_0 = \varepsilon_\infty + \varepsilon_{ion}$, where $\varepsilon_{on}$ accounts for the phonon (ionic) polarization.

The $\varepsilon_{ion}$ term was computed using the PBE functional and density functional perturbation theory (DFPT) at the Γ-point (see SI for details). Since dielectric anisotropy must also be considered, we applied the same capacitor-model correction procedure used for $\varepsilon_\infty$. The final results give $\varepsilon_{0\parallel}^{slab}$ = 4.22 + 20.3 = 24.50 for the 2ML case and $\varepsilon_{0\parallel}^{slab}$ = 4.05 + 21.77 = 25.82 for 5Ml case. Using these values in our mBSE framework — which now includes not only the fast electronic response but also the ionic contribution from low-frequency polar phonons — yields binding energies of $E_b \approx 0.04$ eV for 2ML and $E_b \approx 0.03$ eV for 5ML. Such small values lead to unrealistically low absorption onsets (e.g., in the 2 ML case), in clear disagreement with experiment. This severe underestimation arises from over-screening, as the slow ionic response cannot follow the femtosecond timescale of exciton formation. As emphasized by Laturia *et al.*[28], only the fast electronic component $\varepsilon_\infty$ should be considered when modeling optical excitons in low-dimensional semiconductors. Although the static dielectric constant cannot be directly used in mBSE, it can be interpreted as an approximation to phonon screening, yielding an effective exciton binding energy $E_b^{eff} = E_b^\infty - E_b^{ion}$ (minus sign indicating that $E_b^{eff}$ is smaller than $E_b^\infty$). For 2 ML, this gives $E_b^{eff} = 0.26 - 0.04 = 0.22 \ eV$, and for 5 ML, $E_b^{eff} = 0.21 - 0.03 = 0.18 \ eV$.

The influence of phonon screening on excitons has also been investigated in a broader context. Filip *et. al.*[43] provide a simple analytical estimate of the phonon-screening correction based on the highest LO phonon frequency and the dielectric constants as:

$$\Delta E_b = -2w_{LO}(1 - \frac{\varepsilon_\infty}{\varepsilon_0}) \frac{\sqrt{1 + \frac{w_{LO}}{E_b}} + 3}{(1 + \sqrt{1 + \frac{w_{LO}}{E_b}})^3} \quad (5)$$

Here, $E_b$ denotes the exciton binding energy calculated with purely electronic screening, $w_{LO}$ is the longitudinal optical phonon energy, $\varepsilon_\infty$ and $\varepsilon_0$ are the high-frequency (clamped-ion) and static (relaxed-ion) dielectric constants, respectively. The factor $(1 - \frac{\varepsilon_\infty}{\varepsilon_0})$ accounts for the ionic contribution to dielectric screening, while the rational function involving $\frac{w_{LO}}{E_b}$ captures the interplay between excitonic confinement and phonon dynamics. A negative $\Delta E_b$, therefore, reflects the reduction of the binding energy due to lattice polarization effects. A simplified version, where electron band dispersion is neglected, is also discussed as:

$$\Delta E_b = -2E_b^\infty \frac{\hbar w_{LO}}{\hbar w_{LO} + E_b^\infty}(1 - \frac{\varepsilon_\infty}{\varepsilon_0}) \quad (6)$$

Since the band-resolved contributions and the exciton charge density distributions for the 2ML and 5ML cases (Figure 3) show a dominant contribution to the excitonic wavefunction at the Γ point, it is worth comparing the results from the two approaches. Applying this framework for the 2ML, 3ML, 4ML and 5ML cases (Table 4), here accounting for the anisotropy of the dielectric constant with $\varepsilon_\parallel^{slab}$ and $\varepsilon_{0\parallel}^{slab}$ instead of $\varepsilon_\infty$ and $\varepsilon_0$, yields only a reduction of the excitonic binding energy of 0.01 eV for the full equation and 0.03 eV for 2ML and 5ML in the case of the simplified equation. In any case, both results are only indicative trends and should not be considered as quantitatively accurate values for 2D nanoplatelets, since the Filip et al. formalism was derived for isotropic 3D semiconductors. Rather, they provide a useful upper bound for the strength of phonon-induced renormalization, supporting our conclusion that lattice polarization effects are strongly suppressed by confinement in 2D halide perovskites. This interpretation is consistent with the work of Lee et al.[44], who showed in GaN/AlN quantum wells that phonon corrections decrease monotonically with stronger confinement and saturate beyond the monolayer limit. Our results extend this picture to halide perovskite nanoplatelets, confirming that dimensional reduction strongly mitigates phonon screening and that excitonic properties in 2D perovskites are predominantly governed by the fast electronic dielectric response.

Table 4. Phonon-screened exciton binding energies for $Cs_{(n+1)}Pb_nBr_{3n+1}$ nanoplatelets, estimated using the analytical model[43] based on dielectric constants and LO phonon frequencies.

| n | $w_{LO}$ (eV) | $\varepsilon^{\infty}_{slab,\parallel}$ | $\varepsilon^{slab}_{0\parallel}$ | $E_b$ mBSE (eV) | $\Delta E_b$ (eV) |
|---|---|---|---|---|---|
| 2 | 0.0182 | 4.22 | 20.31 | 0.26 | 0.01 |
| 3 | 0.0177 | 4.11 | 20.79 | 0.24 | 0.01 |
| 4 | 0.0178 | 4.09 | 21.28 | 0.22 | 0.01 |
| 5 | 0.0178 | 4.05 | 21.77 | 0.21 | 0.01 |
| n | $w_{LO}$ (eV) | $\varepsilon^{\infty}_{slab,\parallel}$ | $\varepsilon^{slab}_{0\parallel}$ | $E_b$ mBSE (eV) | $\Delta E_b$ (eV) |
| 2 | 0.0182 | 4.22 | 20.31 | 0.26 | 0.03 |
| 3 | 0.0177 | 4.11 | 20.79 | 0.24 | 0.03 |
| 4 | 0.0178 | 4.09 | 21.28 | 0.22 | 0.03 |
| 5 | 0.0178 | 4.05 | 21.77 | 0.21 | 0.03 |

To further probe vibrational and thermal effects, we employed the phonon-averaged structure obtained from the Zacharias–Giustino (ZG) method which accounts for zero-point motion and finite-temperature displacements through a frozen-phonon geometry for the 2 ML case.[45] Calculations on the ZG-distorted structure reveal a bandgap renormalization of 0.12 eV at 300 K (PBE), reducing the fundamental gap. Within HSE06, the in-plane dielectric constant $\varepsilon^{slab}_{\parallel}$ decreases from 4.22 to 4.09, lowering the exciton binding energy by ~0.03 eV compared to the undistorted case. Importantly, since the bandgap narrowing and the decrease of $E_b$ act in opposite directions, the net shift of the absorption peak is 0.09 eV. While this introduces a shift in the computed binding energies, the change remains within the thermal broadening estimated from Fermi–Dirac statistics near room temperature (about 0.1 eV) and that the robustness of excitonic properties across the 2–5 ML series is largely preserved since we expect trends to hold for thicker nanoplatelets.

Table 5: Dependence of the DDH band gap (Eg), exciton binding energy (Eb) from mBSE on the choice of dielectric constant for nanoplatelets with 2ML and 5ML thickness, and phonon-averaged structure effects at finite temperatures (300 K) from the Zacharias–Giustino (ZG) method

| **2ML** | | | |
|---|---|---|---|
| Dielectric function | Eg (eV) DDH | $E_b$ (eV) mBSE | $E_b$ (eV) 300 K |
| 2.22 $\varepsilon_\perp$ | 4.21 | 0.56 | |
| 3.55 $(2*\varepsilon_\parallel + \varepsilon_\perp)/3$ | 3.35 | 0.29 | |
| 4.22 $\varepsilon_\parallel$ | 3.13 | 0.26 | |

| 5ML | | | |
|---|---|---|---|
| Dielectric function | Eg (eV) DDH | $E_b$ in eV | |
| 2.46 $\varepsilon_\perp$ | 3.70 | 0.39 | |
| 3.52 (2*$\varepsilon_\parallel$ + $\varepsilon_\perp$)/3 | 3.10 | 0.25 | |
| 4.05 $\varepsilon_\parallel$ | 2.88 | 0.21 | |

Our results show that anisotropic dielectric screening is a central factor in shaping the optical and excitonic properties of 2D halide perovskites, rather than a secondary correction. In-plane dielectric components dominate the formation and energetics of optically active excitons, while out-of-plane contributions remain limited in thin-layer regimes (Table 5). This explains the consistent exciton binding energies observed across different thicknesses and provides a robust foundation for extending our framework to a broad range of low-dimensional semiconductors. By combining computational efficiency with a reliable description of confinement trends, our approach is well suited for comparative and high-throughput exploration of layered perovskites. While full GW+BSE remains computationally prohibitive for large supercells with vacuum spacing, the calibrated mBSE method with resolved dielectric constants via the DDH framework offers a practical alternative that can be readily applied to diverse 2D systems at greatly reduced cost.

We emphasize that the originality of our work lies in explicitly coupling dielectric anisotropy using dielectric-dependent functionals with quantum confinement to explain exciton formation in 2D halide perovskite nanoplatelets, revealing effects of the dielectric anisotropy of such a system, benchmarked directly against experiment

Looking ahead, this framework can be extended to assess the influence of temperature, compositional disorder, strain, and external dielectric environments on excitonic responses. For instance, anisotropic 2D excitons have been reported in layered metal–organic chalcogenides, such as silver benzeneselenolate, where the optical response is strongly direction-dependent.[46] Similarly, layered metal–organic chalcogenides have recently been highlighted as a new family of quantum-confined semiconductors with tunable excitonic responses and broad optoelectronic potential.[47] Including such systems would underscore the versatility and transferability of our dielectric-dependent hybrid + mBSE methodology and point toward general design rules for excitonic phenomena across diverse classes of 2D semiconductors. Our work thus provides both a fundamental understanding and a scalable strategy for predictive materials design in nanoscale optoelectronics.

**Supporting Information:**
Supplementary information is available in the online version, including additional computational details, structures used in the calculations and experimental details.

**Code Availability:**
The VASP code is licensed software available from https://www.vasp.at/.

**Acknowledgments:**
We thank Swedish Energy Agency (P2020-90215), FORMAS (Grant no 2022-02297), and the Swedish Research Council (Grant no 2023-05244) for financial support. The computations were performed using resources provided by the National Academic Infrastructure for Supercomputing in Sweden (NAISS) through project NAISS 2024/5-372 and 2025-5-472. S.E.R.-L. acknowledges ANID Fondecyt regular grant number 1220986

**References:**
1. Edvinsson, T. Optical quantum confinement and photocatalytic properties in two-, one- and zerodimensional nanostructures. *R Soc Open Sci* **5**, (2018).
2. Bertolotti, F. *et al.* Crystal Structure, Morphology, and Surface Termination of Cyan-Emissive, Six-Monolayers-Thick CsPbBr3 Nanoplatelets from X-ray Total Scattering. *ACS Nano* **13**, 14294–14307 (2019).
3. Naeem, A. *et al.* Giant exciton oscillator strength and radiatively limited dephasing in two-dimensional platelets. *Phys Rev B Condens Matter Mater Phys* **91**, (2015).
4. Krajewska, C. J. *et al.* A-Site Cation Influence on the Structural and Optical Evolution of Ultrathin Lead Halide Perovskite Nanoplatelets. *ACS Nano* **18**, 8248–8258 (2024).
5. Park, B. W. *et al.* Chemical engineering of methylammonium lead iodide/bromide perovskites: Tuning of opto-electronic properties and photovoltaic performance. *J Mater Chem A Mater* **3**, 21760–21771 (2015).
6. Pazoki, M., Imani, R., Röckert, A. & Edvinsson, T. Electronic structure of 2D hybrid perovskites: Rashba spin-orbit coupling and impact of interlayer spacing. *J Mater Chem A Mater* **10**, 20896–20904 (2022).
7. Pazoki, M. & Edvinsson, T. Metal replacement in perovskite solar cell materials: Chemical bonding effects and optoelectronic properties. *Sustainable Energy and Fuels* vol. 2 1430–1445 Preprint at https://doi.org/10.1039/c8se00143j (2018).
8. Bekenstein, Y., Koscher, B. A., Eaton, S. W., Yang, P. & Alivisatos, A. P. Highly Luminescent Colloidal Nanoplates of Perovskite Cesium Lead Halide and Their Oriented Assemblies. *J Am Chem Soc* **137**, 16008–16011 (2015).
9. Onida, G., Reining, L. & Rubio, A. *Electronic Excitations: Density-Functional versus Many-Body Green's-Function Approaches*.
10. Sun, J., Yang, J. & Ullrich, C. A. Low-cost alternatives to the Bethe-Salpeter equation: Towards simple hybrid functionals for excitonic effects in solids. *Phys Rev Res* **2**, (2020).
11. Tal, A., Liu, P., Kresse, G. & Pasquarello, A. Accurate optical spectra through time-dependent density functional theory based on screening-dependent hybrid functionals. *Phys Rev Res* **2**, (2020).
12. Ketolainen, T., MacHáčová, N. & Karlický, F. Optical Gaps and Excitonic Properties of 2D Materials by Hybrid Time-Dependent Density Functional Theory: Evidences for Monolayers and Prospects for van der Waals Heterostructures. *J Chem Theory Comput* **16**, 5876–5883 (2020).
13. Suzuki, Y. & Watanabe, K. Excitons in two-dimensional atomic layer materials from time-dependent density functional theory: Mono-layer and bi-layer hexagonal boron nitride and transition-metal dichalcogenides. *Physical Chemistry Chemical Physics* **22**, 2908–2916 (2020).
14. Zheng, H., Govoni, M. & Galli, G. Dielectric-dependent hybrid functionals for heterogeneous materials. *Phys Rev Mater* **3**, (2019).
15. Baskurt, M., Erhart, P. & Wiktor, J. Direct, Indirect, and Self-Trapped Excitons in Cs2AgBiBr6. *Journal of Physical Chemistry Letters* **15**, 8549–8554 (2024).

16. Varrassi, L. *et al.* Optical and excitonic properties of transition metal oxide perovskites by the Bethe-Salpeter equation. *Phys Rev Mater* **5**, (2021).
17. Blase, X., Duchemin, I., Jacquemin, D. & Loos, P. F. The Bethe-Salpeter Equation Formalism: From Physics to Chemistry. *Journal of Physical Chemistry Letters* vol. 11 7371–7382 Preprint at https://doi.org/10.1021/acs.jpclett.0c01875 (2020).
18. Cui, Z. H., Wang, Y. C., Zhang, M. Y., Xu, X. & Jiang, H. Doubly Screened Hybrid Functional: An Accurate First-Principles Approach for Both Narrow- and Wide-Gap Semiconductors. *Journal of Physical Chemistry Letters* **9**, 2338–2345 (2018).
19. Chen, W., Miceli, G., Rignanese, G. M. & Pasquarello, A. Nonempirical dielectric-dependent hybrid functional with range separation for semiconductors and insulators. *Phys Rev Mater* **2**, (2018).
20. Liu, P., Franchini, C., Marsman, M. & Kresse, G. Assessing model-dielectric-dependent hybrid functionals on the antiferromagnetic transition-metal monoxides MnO, FeO, CoO, and NiO. *Journal of Physics Condensed Matter* **32**, (2020).
21. Perdew, J. P., Ernzerhof, M. & Burke, K. Rationale for mixing exact exchange with density functional approximations. *Journal of Chemical Physics* **105**, 9982–9985 (1996).
22. Krukau, A. V., Vydrov, O. A., Izmaylov, A. F. & Scuseria, G. E. Influence of the exchange screening parameter on the performance of screened hybrid functionals. *Journal of Chemical Physics* **125**, (2006).
23. Mannino, G. *et al.* Temperature-Dependent Optical Band Gap in CsPbBr3, MAPbBr3, and FAPbBr3 Single Crystals. *Journal of Physical Chemistry Letters* **11**, 2490–2496 (2020).
24. Bechtel, J. S., Thomas, J. C. & Van Der Ven, A. Finite-temperature simulation of anharmonicity and octahedral tilting transitions in halide perovskites. *Phys Rev Mater* **3**, (2019).
25. Hoffman, A. E. J. *et al.* Understanding the phase transition mechanism in the lead halide perovskite CsPbBr3 via theoretical and experimental GIWAXS and Raman spectroscopy. *APL Mater* **11**, (2023).
26. Sio, W. H. & Giustino, F. Unified ab initio description of Fröhlich electron-phonon interactions in two-dimensional and three-dimensional materials. *Phys Rev B* **105**, (2022).
27. Freysoldt, C., Eggert, P., Rinke, P., Schindlmayr, A. & Scheffler, M. Screening in two dimensions: GW calculations for surfaces and thin films using the repeated-slab approach. *Phys Rev B Condens Matter Mater Phys* **77**, (2008).
28. Laturia, A., Van de Put, M. L. & Vandenberghe, W. G. Dielectric properties of hexagonal boron nitride and transition metal dichalcogenides: from monolayer to bulk. *NPJ 2D Mater Appl* **2**, (2018).
29. Bohn, B. J. *et al.* Boosting Tunable Blue Luminescence of Halide Perovskite Nanoplatelets through Postsynthetic Surface Trap Repair. *Nano Lett* **18**, 5231–5238 (2018).
30. Kresse, G. & Furthmü, J. *Efficient Iterative Schemes for Ab Initio Total-Energy Calculations Using a Plane-Wave Basis Set*. (1996).
31. Kresse, G. & Joubert, D. *From Ultrasoft Pseudopotentials to the Projector Augmented-Wave Method*.
32. Perdew, J. P., Burke, K. & Ernzerhof, M. *Generalized Gradient Approximation Made Simple*. (1996).

33. Garba, I. B. *et al.* Three-Dimensional to Layered Halide Perovskites: A Parameter-Free Hybrid Functional Method for Predicting Electronic Band Gaps. *ACS Mater Lett* 1922–1929 (2025) doi:10.1021/acsmaterialslett.5c00158.
34. Akkerman, Q. A. *et al.* Solution Synthesis Approach to Colloidal Cesium Lead Halide Perovskite Nanoplatelets with Monolayer-Level Thickness Control. *J Am Chem Soc* **138**, 1010–1016 (2016).
35. Cucco, B. *et al.* Intrinsic Limits of Charge Carrier Mobilities in Layered Halide Perovskites. *PRX Energy* **3**, (2024).
36. Hansen, K. R. *et al.* Mechanistic origins of excitonic properties in 2D perovskites: Implications for exciton engineering. *Matter* **6**, 3463–3482 (2023).
37. Reyes-Lillo, S. E., Rangel, T., Bruneval, F. & Neaton, J. B. Effects of quantum confinement on excited state properties of SrTiO3 from ab initio many-body perturbation theory. *Phys Rev B* **94**, (2016).
38. Blancon, J. C. *et al.* Scaling law for excitons in 2D perovskite quantum wells. *Nat Commun* **9**, (2018).
39. Miyata, A. *et al.* Direct measurement of the exciton binding energy and effective masses for charge carriers in organic-inorganic tri-halide perovskites. *Nat Phys* **11**, 582–587 (2015).
40. Rasmussen, F. A. & Thygesen, K. S. Computational 2D Materials Database: Electronic Structure of Transition-Metal Dichalcogenides and Oxides. *Journal of Physical Chemistry C* **119**, 13169–13183 (2015).
41. Yang, X. & Zhang, Y. Prediction of high-entropy stabilized solid-solution in multi-component alloys. *Mater Chem Phys* **132**, 233–238 (2012).
42. Jasti, N. P. *et al.* Experimental evidence for defect tolerance in Pb-halide perovskites. *Proc Natl Acad Sci U S A* **121**, (2024).
43. Filip, M. R., Haber, J. B. & Neaton, J. B. Phonon Screening of Excitons in Semiconductors: Halide Perovskites and beyond. *Phys Rev Lett* **127**, (2021).
44. Lee, W. *et al.* Phonon Screening of Excitons in Atomically Thin Semiconductors. *Phys Rev Lett* **133**, (2024).
45. Zacharias, M. & Giustino, F. One-shot calculation of temperature-dependent optical spectra and phonon-induced band-gap renormalization. *Phys Rev B* **94**, (2016).
46. Maserati, L. *et al.* Anisotropic 2D excitons unveiled in organic-inorganic quantum wells. *Mater Horiz* **8**, 197–208 (2021).
47. Paritmongkol, W. *et al.* Layered Metal-Organic Chalcogenides: 2D Optoelectronics in 3D Self-Assembled Semiconductors. *ACS Nano* vol. 19 12467–12477 Preprint at https://doi.org/10.1021/acsnano.4c18493 (2025).